\documentclass[preprint,12pt,authoryear]{elsarticle}

\usepackage{amssymb}
\usepackage{framed}
\usepackage{amsmath,amsfonts}
\usepackage{hyperref}

\usepackage[nameinlink, capitalise]{cleveref}
\Crefname{figure}{Figure}{Figures}
\Crefname{equation}{Equation}{Equations}

\creflabelformat{equation}{#2#1#3}

\usepackage{graphicx}
\usepackage{tabularx}
\usepackage{booktabs}
\usepackage{multirow}
\usepackage{bm}
\usepackage{mathtools}
\usepackage{siunitx}
\usepackage{caption}
\usepackage{subcaption}
\usepackage{float}
\usepackage{comment}
\usepackage{placeins}
\usepackage{enumitem}
\usepackage[export]{adjustbox}

\begin{document}

\begin{frontmatter}



\title{Parallel Gaussian Process with Kernel Approximation in CUDA}


\author[label1]{Davide Carminati}

\affiliation[label1]{organization={Department of Mechanical and Aerospace Engineering, Politecnico di Torino},%Department and Organization
            addressline={Corso Duca degli Abruzzi 24}, 
            city={Torino},
            postcode={10129}, 
            country={Italy}}

\begin{abstract}
This paper introduces a parallel implementation in CUDA/C++ of the Gaussian process with a decomposed kernel. This recent formulation, introduced by \citet{fastapproxGP}, is characterized by an approximated -- but much smaller -- matrix to be inverted compared to plain Gaussian process. However, it exhibits a limitation when dealing with higher-dimensional samples which degrades execution times. The solution presented in this paper relies on parallelizing the computation of the predictive posterior statistics on a GPU using CUDA and its libraries. The CPU code and GPU code are then benchmarked on different CPU-GPU configurations to show the benefits of the parallel implementation on GPU over the CPU. 

\end{abstract}


\end{frontmatter}


\section{Introduction} \label{sec:intro}
Gaussian process (GP) is a well-established method in Machine Learning. Its applications range from economics, medicine and geophysics to engineering and data science. In its most used formulation for regression problems, the prior and the likelihood are considered as Gaussian distributions. This allows to find an analytical solution for the predictive posterior in terms of mean and covariance functions. When dealing with huge datasets, the computation of the predictive posterior could be infeasible due to huge matrix inversion. Different approaches have been proposed to mitigate the problem, among which the most known is the sparse formulation \citep{quinonero2005unifying,snelson2005sparse,titsias2009variational,pmlr-v51-matthews16}. The common aspect among all sparse GP approaches is that only a small subset of the latent variables are treated exactly, and whose locations can be determined using different strategies. Selecting the most suitable strategy is in fact the main challange of this approach. Several sparse GP formulations are available as open-source libraries, \textit{e.g.}, \textit{GPflow} \citep{GPflow2020multioutput}, which is based on \textit{Tensorflow}; \textit{GPyTorch} \citep{gardner2021gpytorch}, implemented in \textit{PyTorch}; \textit{GPJax} \citep{Pinder2022}, based on \textit{JAX}; and \textit{GPy} \citep{gpy2014}. \\
% \citep{tensorflow2015-whitepaper}
% \citep{paszke2019pytorch}
% \citep{jax2018github}
Kernel approximation techniques are an alternative approach for approximating the prior, allowing to process huge datasets. Several kernel approximation techniques can be found in the literature \citep[e.g.][]{seeger2000,peng2015}. More recently, a formulation relying on the spectral decomposition of the kernel has been proposed by \citet{fastapproxGP}. By using Mercer's theorem, it is possible to express the kernel as the product between its eigenvalues and eigenvectors, and obtain a predictive posterior in which the matrix to be inverted is as big as the number of considered eigenvalues. This allows to have a trade-off between accuracy of the solution and execution speed. The authors show high accuracy with a small number of eigenvalues and linear execution time with respect to the number of samples. However, while the original paper addresses the problem of multi-output Gaussian process, it does not cover the case in which the train samples are multidimensional. The extension of the kernel decomposition for higher sample dimensions is treated here, but it has a significant shortcoming. In fact, 
the number of combinations of eigenvalues grows exponentially with the number of sample dimensions, leading to huge matrix sizes even for small numbers of eigenvalues. 
The solution proposed in this paper relies on parallelizing the computation of the predictive posterior statistics on a Graphics Processing Unit (GPU), exploiting the general matrix-matrix multiplication (GEMM) routine based on the BLAS (Basic Linear Algebra Subprograms) library.  This results in lower execution times with respect to the CPU implementation when sample dimensions grow, preserving the benefits of the formulation for more general problems.\\
The paper is organized as follows. In \autoref{sec:theory}, the mathematical background regarding Gaussian processes and kernel decomposition is presented. \autoref{sec:3} describes the limitation of the approach and the proposed implementations to be compared. Then, \autoref{sec:results} analyzes the obtained results using the proposed parallel implementation.

\section{Theoretical background} \label{sec:theory}
\subsection{Gaussian process regression}
Consider a training dataset $(X, \mathbf{y})$, where $X = \{\mathbf{x}_i \in \mathbb{R}^p \, | \, i = 1,\ldots , N\}$ and $\mathbf{y} = \{ y_i \in \mathbb{R} \, | \, i = 1,\ldots , N\}$, and a test dataset $(X_*, \mathbf{y}_*)$, where $X = \{\mathbf{x}_{*i} \in \mathbb{R}^p \, | \, i = 1,\ldots , N_*\}$ and $\mathbf{y}_* = \{ y_{*i} \in \mathbb{R} \, | \, i = 1,\ldots , N_*\}$. Also, consider the following regression model:
\begin{equation}
    y = f(\mathbf{x}) + \varepsilon
\end{equation}
where $\varepsilon \sim \mathcal{N}(0, \, \sigma^2)$ represent the white noise affecting the model. A Gaussian distribution $p(\mathbf{f})$ is imposed as prior, i.e. $p(\mathbf{f}) = \mathcal{N}(\mathbf{0}, \, k(X,X))$, where $k(\cdot, \cdot)$ is the kernel and $\mathbf{f} = f(\mathbf{x})$ for shorthand. The predictive posterior over the test dataset writes:
\begin{equation}
	\mathbf{f}_* | \mathbf{f}, \mathbf{y}, \bm{\theta} \sim \mathcal{N}(\bm{\mu}_*, \Sigma_*)
\end{equation}
where:
\begin{align}
	\bm{\mu}_* &= m(X_*) + K_* (K + \sigma^2 I)^{-1} (\mathbf{y} - m(X))\label{eq:muGPposterior} \\
	\Sigma_* &= K_{**} - K_* (K + \sigma^2 I)^{-1} K_*^T \label{eq:sigmaGPposterior}
\end{align}
in which $K = k(X, X)$ is a $N \times N$ matrix, $K_* = k(X_*, X)$ is a $N_* \times N$ matrix, $K_{**} = k(X_*, X_*)$ matrix, $\bm{\mu}_*$ is a $N_* \times 1$ vector, while $\Sigma_*$ is a $N_* \times N_*$ matrix.  In most of the applications, the zero-mean GP is employed by setting $m(X) = m(X_*) = \mathbf{0}$. 
% Note that solving \cref{eq:muGPposterior,eq:sigmaGPposterior} means inverting the matrix $(K + \sigma_n^2 I)$, which involves $\mathcal{O}(N^3)$ operations.
\subsection{Gaussian process regression with a decomposed kernel}
When dealing with huge datasets, the major issue is essentially the inversion of the kernel $\tilde{K} = (K + \sigma^2 I)$ involving $\mathcal{O}(N^3)$ operations. A direct solution to this problem is exploiting a common property to all kernels which follows from \textit{Mercer's theorem}. The kernel is approximated with the sum of the product between its eigenvalues and its eigenvectors, and it is possible to achieve a more lightweight inverse computation. This approach has been originally developed by \cite{fastapproxGP}. We report the mathematical derivation of the GP predictive posterior when considering a decomposed kernel, following the computation presented in the original paper. \\
The decomposition is possible due to \textit{Mercer's theorem} (see \textit{e.g.}, \cite{steinwart_mercers_2012}), which states that:
\begin{equation}
    k(\mathbf{x}, \mathbf{x'}) = \sum_{i = 1}^{\infty} \lambda_i \phi_i(\mathbf{x}) \phi_i(\mathbf{x'})
\end{equation}
where $\lambda_i$ and $\phi_i(\mathbf{x})$ are the $i$-th kernel eigenvalue and kernel eigenfunction respectively.
Using a finite series approximation of $n \ll N$ eigenvalues, one obtains:
\begin{equation}
	k(\mathbf{x}, \mathbf{x'}) \approx \sum_{i = 1}^{n} \lambda_i \phi_i(\mathbf{x}) \phi_i(\mathbf{x'})
\end{equation}
which can be rearranged as a matrix product:
\begin{equation}\label{eq:approx_kernel}
	k(X,X') \approx \bm{\Phi}_{(X)} \Lambda \bm{\Phi}^T_{(X')}
\end{equation}
where $\bm{\Phi}_{(X)} = \begin{bsmallmatrix}
	| & & | \\
	\phi_i(X) & \ldots & \phi_n(X)\\
	| & & |
\end{bsmallmatrix} $ and $\Lambda = \begin{bsmallmatrix}
	\lambda_1 & & \\
	& \ddots & \\
	& & \lambda_n
\end{bsmallmatrix}$. Substituting this result into \cref{eq:muGPposterior,eq:sigmaGPposterior}, one obtains:
\begin{align}
	\bm{\mu}_* &\approx m(X_*) + \bm{\Phi}_{(X_*)} \Lambda \bm{\Phi}^T_{(X)} \left( \bm{\Phi}_{(X)} \Lambda \bm{\Phi}^T_{(X)} + \Sigma_n \right)^{-1} (\mathbf{y} - m(X)) \label{eq:muapproxGP} \\
	\Sigma_* &\approx \bm{\Phi}_{(X_*)} \Lambda \bm{\Phi}^T_{(X_*)} - \nonumber \\
	& \hspace{15pt} \bm{\Phi}_{(X_*)} \Lambda \bm{\Phi}^T_{(X)} \left( \bm{\Phi}_{(X)} \Lambda \bm{\Phi}^T_{(X)} + \sigma_n^2 I \right)^{-1} \bm{\Phi}_{(X)} \Lambda \bm{\Phi}^T_{(X_*)} \label{eq:sigmaapproxGP}
\end{align}
where $\Sigma_n = \sigma_n^2 I$ is the noise covariance in a more compact notation. Applying the Woodbury identity, the inverse of the approximated kernel can be rewritten as:
\begin{equation}\label{eq:woodburyGP}
	\left( \bm{\Phi}_{(X)} \Lambda \bm{\Phi}^T_{(X)} + \Sigma_n \right)^{-1} =
	\Sigma_n^{-1} - \Sigma_n^{-1} \bm{\Phi}_{(X)} \left( \Lambda^{-1} + \bm{\Phi}^T_{(X)} \Sigma_n^{-1} \bm{\Phi}_{(X)} \right)^{-1} \bm{\Phi}^T_{(X)} \Sigma_n^{-1}
\end{equation}
In this way, only a smaller $n\times n$ matrix $ \bar{\Lambda} = \left( \Lambda^{-1} + \bm{\Phi}^T_{(X)} \Sigma_n^{-1} \bm{\Phi}_{(X)} \right)$ has to be inverted, involving $\mathcal{O}(n^3)$ operations.
% thus allowing a much lower computational cost when computing the GP prediction. 
Moreover, the terms $\Lambda^{-1}$ and $\Sigma_n^{-1}$ are unproblematic as they are diagonal matrices. By substituting \autoref{eq:woodburyGP} into \cref{eq:muapproxGP,eq:sigmaapproxGP}, one obtains the Fast-Approximate Gaussian Process (FAGP) formulation:

\begin{align}
    \bm{\mu}_* &\approx m(X_*) + W (\mathbf{y} - m(X)) \label{eq:muFAGP} \\    
    \Sigma_* &\approx \mathbf{\Phi}_{(X_*)} \Lambda \mathbf{\Phi}_{(X_*)}^T - W \mathbf{\Phi}_{(X)} \Lambda \mathbf{\Phi}_{(X_*)}^T \label{eq:sigmaFAGP}
\end{align}
in which
\begin{align*}
	W &= \mathbf{\Phi}_{(X_*)} \Lambda \mathbf{\Phi}_{(X)}^T \left( \Sigma_N^{-1} - \Sigma_N^{-1} \mathbf{\Phi}_{(X)} \bar{\Lambda}^{-1} \mathbf{\Phi}_{(X)}^T \Sigma_N^{-1} \right) \\
	\bar{\Lambda} &= \Lambda^{-1} + \mathbf{\Phi}_{(X)}^T \Sigma_N^{-1} \mathbf{\Phi}_{(X)}
\end{align*}
\subsection{Kernel decomposition}
While there are a good number of Mercer expansions for the most used kernels, \textit{e.g.} Matérn, periodic, etc., in this work only the widely used square exponential (SE) kernel is considered. We report below the analytical expression of the eigenfunctions $\bm{\Phi}_{(\cdot)}$ and eigenvalues $\lambda_i$ for the square exponential kernel. To ease the notation, consider the univariate square exponential kernel with unitary amplitude:
% and set $\varepsilon = \frac{1}{\sqrt{2}l}$:
\begin{equation}
	k_{SE}(x,x') = \exp \left( -\varepsilon^2 (x -x')^2 \right)
\end{equation}
Following \cite{fasshauer2012stable}, we set:
\small
\begin{align}
	\beta &= \left(1 + \left(\frac{2\varepsilon}{\rho}\right)^2\right)^\frac{1}{4}, && \gamma_i = \sqrt{\frac{\beta}{2^{i-1}\Gamma_{(i)}}}, && \delta^2 = \frac{\rho}{2}(\beta^2-1),
\end{align}
\normalsize
where $\rho$ is the \textit{global} scale factor. The square exponential kernel Mercer expansion leads to the following eigenfunctions:
% \cite{fasshauer2011positive}:
\begin{align}\label{eq:eigfun}
	\phi_i(x) = \gamma_i \exp^{-\delta^2 x^2} H_{i-1}(\rho \beta x) && i = 1, \ldots, n
\end{align}
where $H_{i-1}$ is the \textit{classical} Hermite polynomial of degree $i-1$.
Their corresponding eigenvalues are defined as:
\begin{align}\label{eq:eigenvalues}
	\lambda_i = \sqrt{\frac{\rho^2}{\rho^2 + \delta^2 + \varepsilon^2}} \left(\frac{\varepsilon^2}{\rho^2 + \delta^2 + \varepsilon^2} \right)^{i-1}&& i = 1, \ldots, n
\end{align}
In the decomposition, there is one additional parameter $\rho$, which controls how fast the eigenvalues decrease. \\
In the original paper, only the unidimensional case is covered. Now, we generalize the kernel decomposition to a multidimensional case. Consider the multivariate square exponential kernel with Automatic Relevance Determination (ARD):
\begin{equation}
	k_{SE}^{ARD}(\mathbf{x}, \mathbf{x'}) = \exp \left( -\varepsilon_1^2 (x_1 - x_1')^2 - \ldots - \varepsilon_p^2 (x_p - x_p')^2 \right)
\end{equation}
where $\mathbf{x} = \begin{bsmallmatrix}
	x_1 & x_2 & \ldots & x_p
\end{bsmallmatrix} \in \mathbb{R}^p$. For $p$-variate kernels, it holds that:
\begin{equation}
k_{SE}^{ARD}{(\mathbf{x},\mathbf{x'})} \approx  \sum_{\mathbf{n}\in \mathbb{N}^p} \lambda_{\mathbf{n}} \phi_{\mathbf{n}}(\mathbf{x}) \phi_{\mathbf{n}}(\mathbf{x'})
\end{equation}
where $\mathbf{n}$ is the set of all $n^p$ combination of the considered number of eigenvalues.
The eigenvalues $\lambda_{\mathbf{n}}$ and eigenfunctions $\phi_{\mathbf{n}}(\mathbf{x})$ are defined as in \cite{fasshauer2012stable}:
\begin{align}
\phi_{\mathbf{n}}(\mathbf{x}) &= \prod_{j=1}^{p} \phi_{n_{j}}(x_j; \varepsilon_j, \rho_j) \\
\lambda_{\mathbf{n}} &= \prod_{j=1}^{p} \lambda_{n_j} (\varepsilon_j, \rho_j)
\end{align}
where $\phi_{n_{j}}(x_j; \varepsilon_j, \rho_j)$ and $\lambda_{n_j}(\varepsilon_j, \rho_j)$ are defined in \autoref{eq:eigfun} and \autoref{eq:eigenvalues} respectively using the length scale $\varepsilon_j$ and global scale factor $\rho_j$ relative to the $j$-th sample dimension.\\
We now expose the limitation of this formulation arising when dealing with higher-dimension problems, which degrades the execution time.\\
Computing \cref{eq:muFAGP,eq:sigmaFAGP} involves a higher number of (generally dense) matrix-matrix multiplications compared to \cref{eq:muGPposterior,eq:sigmaGPposterior}. Also, for multidimensional cases, both $\mathbf{\Phi}_{(X)}$ and $\Lambda$ in \autoref{eq:approx_kernel} are subject to a massive increase of their size when considering more eigenvalues. In fact, $\Lambda$ is a $n^p \times n^p$ matrix while $\mathbf{\Phi}_{(X)}$ is a $N \times n^p$ matrix. To cope with these potentially huge matrices, a parallel implementation on a CUDA device is explored and compared with its CPU counterpart.

\section{Implementation details}\label{sec:3}
The CPU implementation of this formulation is done in \verb|C++| using \textit{Eigen}, which is one of the most used linear algebra libraries available. \textit{Eigen} employs \textit{OpenMP} for parallelization on CPU when performing the GEMM routine\footnote{See \url{https://eigen.tuxfamily.org/dox/TopicMultiThreading.html} for a complete list of multithreaded operations.}.\\
The GPU implementation is written in \verb|CUDA/C++| and exploits \textit{cuBLAS} and \textit{cuSOLVER}, which are high level libraries provided by NVIDIA that allow to write host code functions for the device. \textit{cuBLAS} provides BLAS-like functions for dense matrix-vector and dense matrix-matrix multiplication. \textit{cuSOLVER} provides routines for LU factorization and linear system solve which are used for inverting $\bar{\Lambda}$. 
% Finally, \textit{cuSPARSE} efficiently handles the multiplication of $\Lambda \mathbf{\Phi}_{(X)}^T = \left( \mathbf{\Phi}_{(X)} \Lambda \right)^T$ in \cref{eq:muFAGP,eq:sigmaFAGP}. 
Implementations in \verb|OpenGL| and \verb|ROCm| are also possible when no NVIDIA GPU is available. \\
A bash script automatically creates some example train datasets with increasing number of eigenvalues $n$ and increasing sample dimension $p$ while keeping the number of samples $N$ constant and equal to 10000. This allows to assess only the effect that $n$ and $p$ have on the execution time. The function used to generate the samples is:
\begin{equation}
    y = \sum_{i=1}^p \cos{x_i} + \nu
\end{equation}
where $\nu \sim \mathcal{N}(0, \, \sigma_n^2)$ is a white noise with known variance corrupting the observations. The execution times are computed carrying out Monte Carlo simulations for all the combinations of considered number of eigenvalues and sample dimensions\footnote{The code used to generate the datasets and the results contained in this paper is available at \url{https://github.com/DavideCarminati/cuFAGP}.}.
Tests are conducted on three different machines to prove the effectiveness of the implementation even on low-range devices. Their main characteristics are reported in \autoref{table:conf}.
\begin{table}[h]
    \centering
    \begin{tabular}{cccc}
        \toprule
        \multirow{2}{*}{} & \multicolumn{3}{c}{Configuration} \\
        \cmidrule{2-4}
        & A & B & C\\
        \midrule
        Host CPU & Intel i7-7700 & AMD Ryzen 5 3600 & Intel i7-7700HQ\\
        Host RAM & 32GB & 32GB & 16GB\\
        Nvidia GPU & Quadro P2200 & RTX 2080 Super & GTX 1050 Mobile\\
        GPU RAM & 5GB GDDR5X & 8GB GDDR6 & 4GB GDDR5\\
        % OS & \multicolumn{3}{c}{Ubuntu 20.04.6 LTS}\\
        \bottomrule
    \end{tabular}
    \caption{Hardware of the computers used for tests.}\label{table:conf}
\end{table}
The operations timed during execution are the following:
\begin{itemize}[noitemsep]
    \item memory allocation and memory copies from host to device (GPU only),
    \item computation of eigenvalues and eigenfunctions, 
    \item computation of the posterior predictive mean, and
    \item memory copy from device to host (GPU only).
\end{itemize}

\section{Results and discussion} \label{sec:results}
\begin{figure}[ht]
	\centering
    \hspace{3mm}
	\begin{subfigure}[b]{0.45\textwidth}
        \centering
		% \makebox[\textwidth]{
		\includegraphics[width=\textwidth]{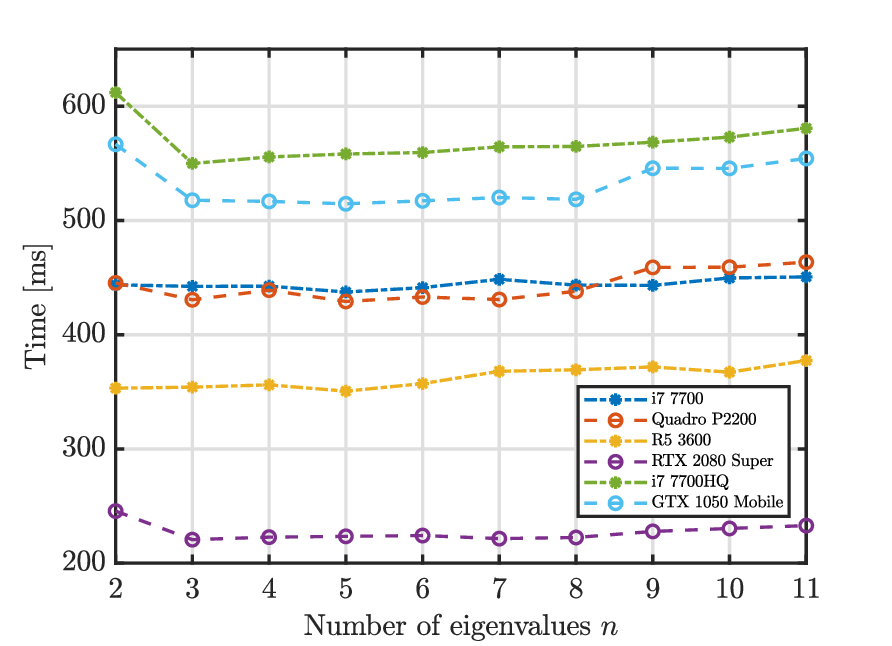}
		% }
		\caption{Dimension $p = 1$}
        \label{fig:1D}
	\end{subfigure}
    \hspace{4mm}
    \centering
	\begin{subfigure}[b]{0.45\textwidth}
		\centering
        % \makebox[\textwidth]{
			\includegraphics[width=\textwidth]{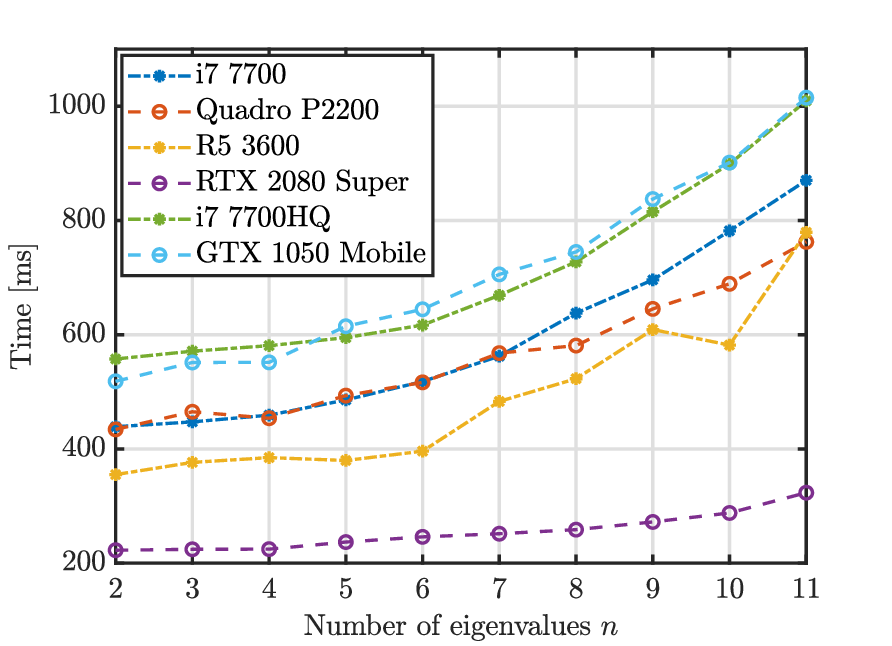}
		% }
		\caption{Dimension $p = 2$}\label{fig:2D}
	\end{subfigure}
    \hspace{2mm}
    \centering
	\begin{subfigure}[h!]{0.45\textwidth}
		\centering
        \makebox[\textwidth]{
			\includegraphics[width=\textwidth]{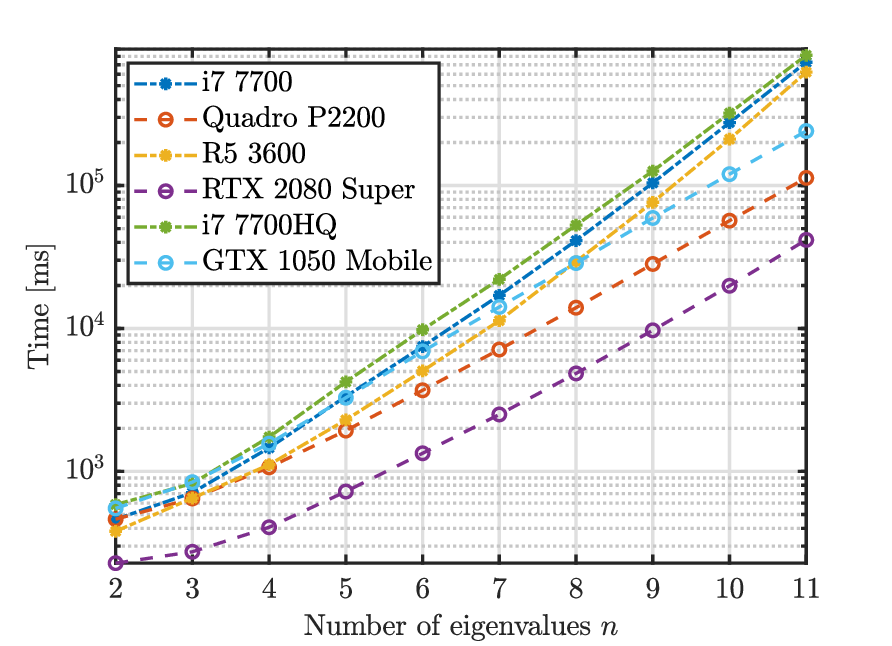}
		}
		\caption{Dimension $p = 4$}\label{fig:4D}
	\end{subfigure}
    \caption{Execution times when varying the number of dimensions $p$ and eigenvalues $n$}\label{fig:fagp}
\end{figure}

Tests are carried out for an increasing number of considered eigenvalues and for sample dimensions $p = 1$, $2$ and $4$. \autoref{fig:fagp} shows the execution time of both the implementations. When $p=1$, the GPU has no particular advantage over the CPU, proving that the parallel implementation can be used as well for unidimensional samples without degradation in the execution speed. However, more powerful GPUs can gain a small margin over the CPU. When $p = 2$, the CPU struggles with a higher number of eigenvalues $n$, as matrix sizes increase with the power of two (\textit{e.g.}, if $n = 9$ then $\Lambda$ is a $9^2 \times 9^2$ matrix, and $\Phi$ a $N \times 9^2$ matrix). Finally, when $p = 4$ all GPUs -- regardless of their power -- are able to outrun their respective CPUs when $n$ increases. The high-range RTX 2080 Super can be roughly 15 times faster the Ryzen 5 CPU, taking only $\approx 40$ s instead of $\approx 10.5$ min for $n = 11$.\\
The benchmark comparison shows that parallelizing on GPU is beneficial to this GP formulation, even when using low-range or mobile GPUs. More up-to-date and powerful GPUs should be able to outrun the CPU even with small $p$ and $n$. 

\section{Conclusion}\label{sec:conclusion}
In this paper, we show a limitation of the kernel approximation method introduced by \citet{fastapproxGP} when dealing with higher dimensional samples. We generalize the original formulation for multidimensional samples and propose a parallel implementation in CUDA for computing the GP predictive posterior. The results show that the GPU implementation scales better with the number of considered eigenvalues and sample dimensions than the CPU counterpart. In this way, it is possible to generalize the original formulation for multidimensional problems while preserving performance in terms of execution speed, even for low-range or mobile GPUs.\\
As a final note, in the original paper the kernel approximation is used also for hyperparameter learning. A parallel implementation of the optimization problem for hyperparameter learning is currently in development and will be a future work.

\section*{Acknowledgement}
The author would like to thank Prof. Elisa Capello and Prof. Stefano Primatesta for the equipment and for all the feedback, and Enza Trombetta and Iris David Du Mutel for the help during the implementation and for reviewing the paper.

\appendix

\section{Computational details}
The results in this paper were obtained using Ubuntu 20.04.6 LTS. The \verb|C++| compiler is \verb|gcc| version 9.4.0. The \verb|CUDA| toolkit version is 12.2.140, while the NVIDIA driver version is 535.129.03-server. The non-server driver version achieves worse results due to the GPU executing processes related to the desktop environment.\\
To build the CPU/GPU binaries, \verb|CMake| $\geq 3.18$ and \textit{Eigen} 3.4.0 are needed.\\
The figures are generated in \verb|MATLAB| R2022a. The \verb|MATLAB| scripts are also tested in \verb|Octave| 5.2.0.

\bibliographystyle{elsarticle-harv} 
\bibliography{refs}





% \end{thebibliography}

\end{document}